\documentclass[preprint]{aastex}
\usepackage{lscape} 
\usepackage{amsmath, amsthm,amssymb}
\usepackage{multirow}
\usepackage{url}

\title{Limits on Quaoar's Atmosphere}

\author{Wesley C. Fraser {$^1$}}
\altaffiltext{1}{Herzberg Institute of Astrophysics, 5071 W. Saanich Rd. Victoria, BCV9E 2E7}
\email{wesley.fraser@nrc.ca}
\author{Chad Trujillo {$^2$}}
\author{Andrew W. Stephens {$^2$}}
\author{German Gimeno {$^3$}}
\author{Michael E. Brown {$^4$}}
\author{Stephen Gwyn {$^1$}}
\author{JJ Kavelaars {$^1$}}

\altaffiltext{2}{Gemini Observatory, Northern Operations Center, 670 N A'ohoku Place, Hilo, HI 96720}
\altaffiltext{3}{Gemini Observatory, Southern Operations Center, c/o AURA, Casilla 603, La Serena, Chile}
\altaffiltext{4}{California Institute of Technology, Division of Geological and Planetary Sciences, 1200 E. California Blvd., Pasadena, CA. 91101 USA}

\date{} 

\slugcomment{Submitted to the Astrophysical Journal Letters}
\received{É}
\revised{É}

\begin{abstract}
Here we present high cadence photometry  taken by the Acquisition Camera on Gemini South, of a close passage by the $\sim540$~km radius Kuiper Belt Object, (50000) Quaoar, of a r'=20.2 background star. Observations before and after the event show that the apparent impact parameter of the event was $0.019\pm0.004$", corresponding to a close approach of $580\pm120$~km to the centre of Quaoar. No signatures of occultation by either Quaoar's limb or its potential atmosphere are detectable in the relative photometry of Quaoar and the target star, which were unresolved during closest approach. From this photometry we are able to put constraints on any potential atmosphere Quaoar might have. Using a Markov chain Monte Carlo and likelihood approach, we place pressure upper limits on sublimation supported, isothermal atmospheres of pure N$_2$, CO, and CH$_4$. For N$_2$ and CO, the upper limit surface pressures are 1 and 0.7 $\mu\mbox{bar}$ respectively. The surface temperature required for such low sublimation pressures is $\sim33$~K, much lower than Quaoar's mean temperature of $\sim44$~K  measured by others. We conclude that Quaoar cannot have an isothermal N$_2$ or CO atmosphere. We cannot eliminate the possibility of a CH$_4$ atmosphere, but place upper surface pressure and mean temperature limits of $\sim138$~nbar and $\sim44$~K respectively.
\end{abstract}

\begin{document}

\maketitle

\section{Introduction \label{sec:Intro}}
Barring in-situ observations by satellite, observation of a stellar occultations is one of the only techniques by which the characterization of Kuiper Belt Object (KBO) atmospheres is possible. The atmosphere of Pluto was discovered by occultations \citep{Hubbard1988, Elliot1989}. Further, the atmosphere of Triton has been characterized in a similar fashion by \citet[see, for example][]{Elliot1998}. It is now generally accepted that the atmospheres of Triton and Pluto consist primarily of N$_2$, CO, and CH$_4$ and are supported by sublimation of those ices from their surfaces \citep[for reviews on the subject, see][]{Trafton1997,Yelle1997}. 

Some of the same ices that support Pluto's atmosphere have been detected (or inferred) on other Kuiper Belt Objects:
Eris is known to posseses CH$_4$ and possibly has N$_2$ ice on its surface \citep{Tegler2010};
Makemake and Quaoar both posses CH$_4$ \citep{Brown2007c,Schaller2007b,Tegler2010}; 
and tentative detections have been made of  CH$_4$ on (225088) 2007 OR10 and both CH$_4$ and N$_2$ on Sedna \citep{Brown2011b,Barucci2005}.
The surface temperatures of these objects are $\sim35-55$~K \citep{Stansberry2008}, similar to that of Pluto. It is expected that, like Pluto, these objects posses atmospheres with surface pressures at the nanobar (for CH$_4$) to microbar (for N$_2$) level. 

We present observations of a stellar appulse event that probed the region tens to hundreds of kilometres above the surface of the KBO Quaoar. From these observations we can formally eliminate the possibility that Quaoar possesses an N$_2$ atmosphere, and place strong upper limits on the pressure of a CH$_4$ dominated atmosphere. In Section 2, we present the observations of the event and our data reductions. In Section 3, we place upper limits on the surface pressure of Quaoar's atmosphere, and we finish with concluding remarks in Section 4.

\section{Observations and Reductions \label{sec:observations}}
Quaoar was predicted to have a close passage with a star at RA and Dec. of 17:28:58.02 and -15:22:51.19 (J2000) with brightness r'=20.2 on July 13th (Fraser et al. in press). The prediction placed Quaoar's shadow trajectory across South America near the Gemini South (GS) telescope. At Gemini South, the time of occultation centre was predicted to occur at 04:32 UT with an apparent impact parameter of $0.017\pm0.03$" and a ground shadow velocity of $21.4 \mbox{ km s$^{-1}$}$. Observations from the Gemini Multi-Object Spectrograph (GMOS) on GS taken roughly 2 hours before and after the event (program GS-2012A-DD-4) reveal that the closest approach was $0.019\pm0.004$", corresponding to a close approach of $580\pm120$~km at a time of  04:25:52$\pm7$~s UT defined by the internal Gemini system time (Fraser et al., In Press).

\begin{figure}[h]
   \centering
   \plotone{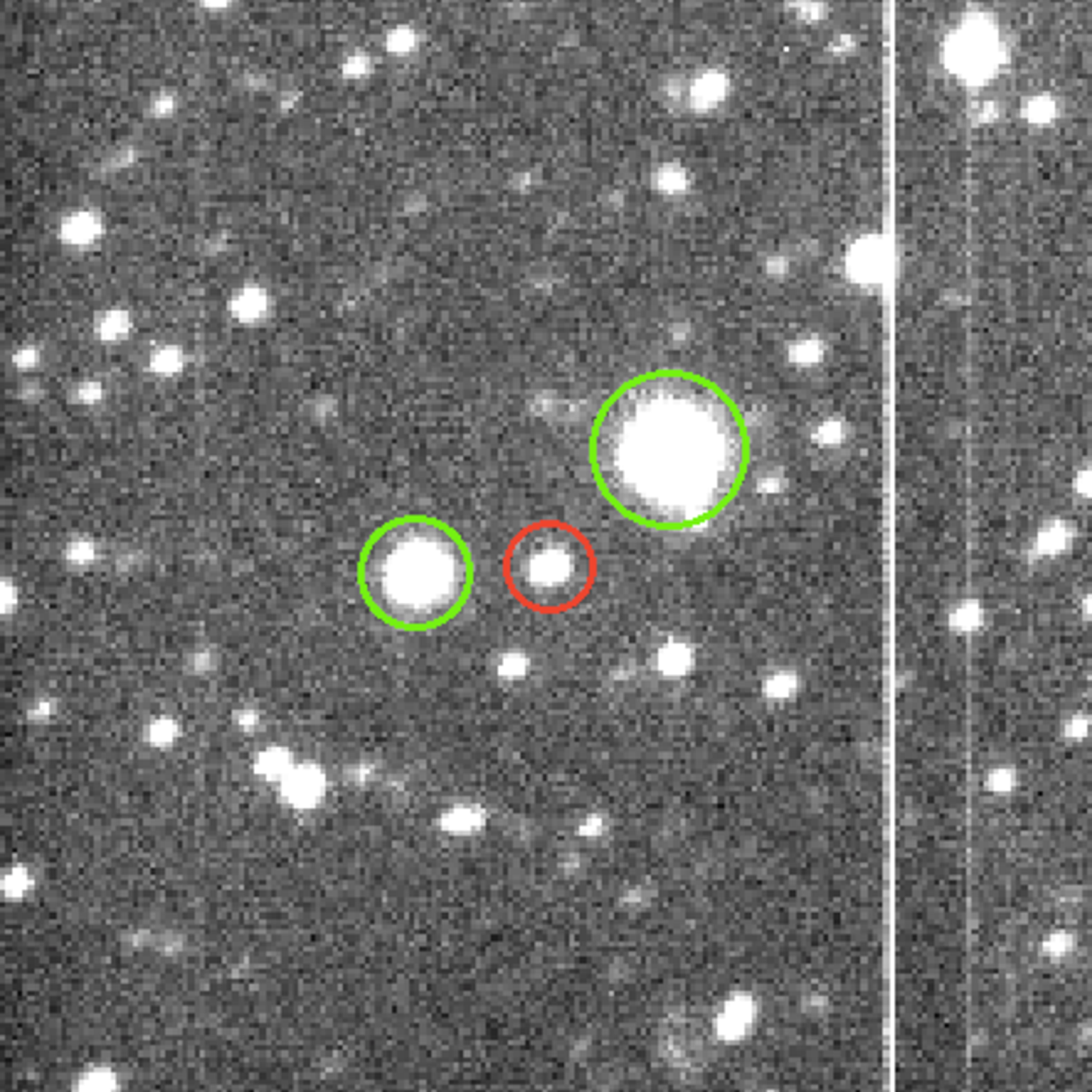} 
  \figcaption{Example Acquisition Camera image. The reference stars are marked by the large green circles. The elongated Quaoar+target star is marked in the centre of the image by the small red circle.  \label{fig:AC_image}}
\end{figure}

Observations of the event itself were made with the Acquisition Camera (AC) on GS. The target star was observed with a cosmetically clean 200x200 pixel section of the camera in a 20 minute window centred on the predicted closest approach time and consisted of 1~s exposures, resulting in a median imaging frequency of 0.7~Hz. The observations were made to include two bright reference stars, brightness r'=15.8 and r'=17.8, in the images. An example image is shown in Figure~\ref{fig:AC_image}.

Due to limitations in the AC control software, the beginning and end times of each exposure were recorded with only integer second precision. We found that the exposure times could be adequately corrected to sub-integer second precision as follows. Starting from the integer accuracy exposure start times contained in the AC image headers, $t_{\textrm{i,H}}$, a mean imaging period, $P=1.44$~s was determined. New start times for the ith exposure were then estimated as $t_{\textrm{i,E}}=P~i+t_{\textrm{o,H}}$.  A running mean of time difference between the header recorded time and the time estimate from the mean period, $\Delta t=t_{\textrm{i,E}}-t_{\textrm{H}}$ was taken every 9 images. Assuming the imaging frequency, or $\Delta t$, smoothly varied with time, cubic spline interpolation $S(t)$, of $\Delta t$ vs. $t_{\textrm{i,E}}$, was then used to improve the estimated time $t_{\textrm{i,E}}$; the adopted exposure start times are, $t_\textrm{i}=t_{\textrm{i,E}}+S(t_{\textrm{i,E}})$. This method resulted in corrected times that were less than half the mean image period away from the start times, $t_{\textrm{i,H}}$, exactly as expected if the start times were rounded to integer precision. 

\begin{figure}[h]
   \centering
   \plotone{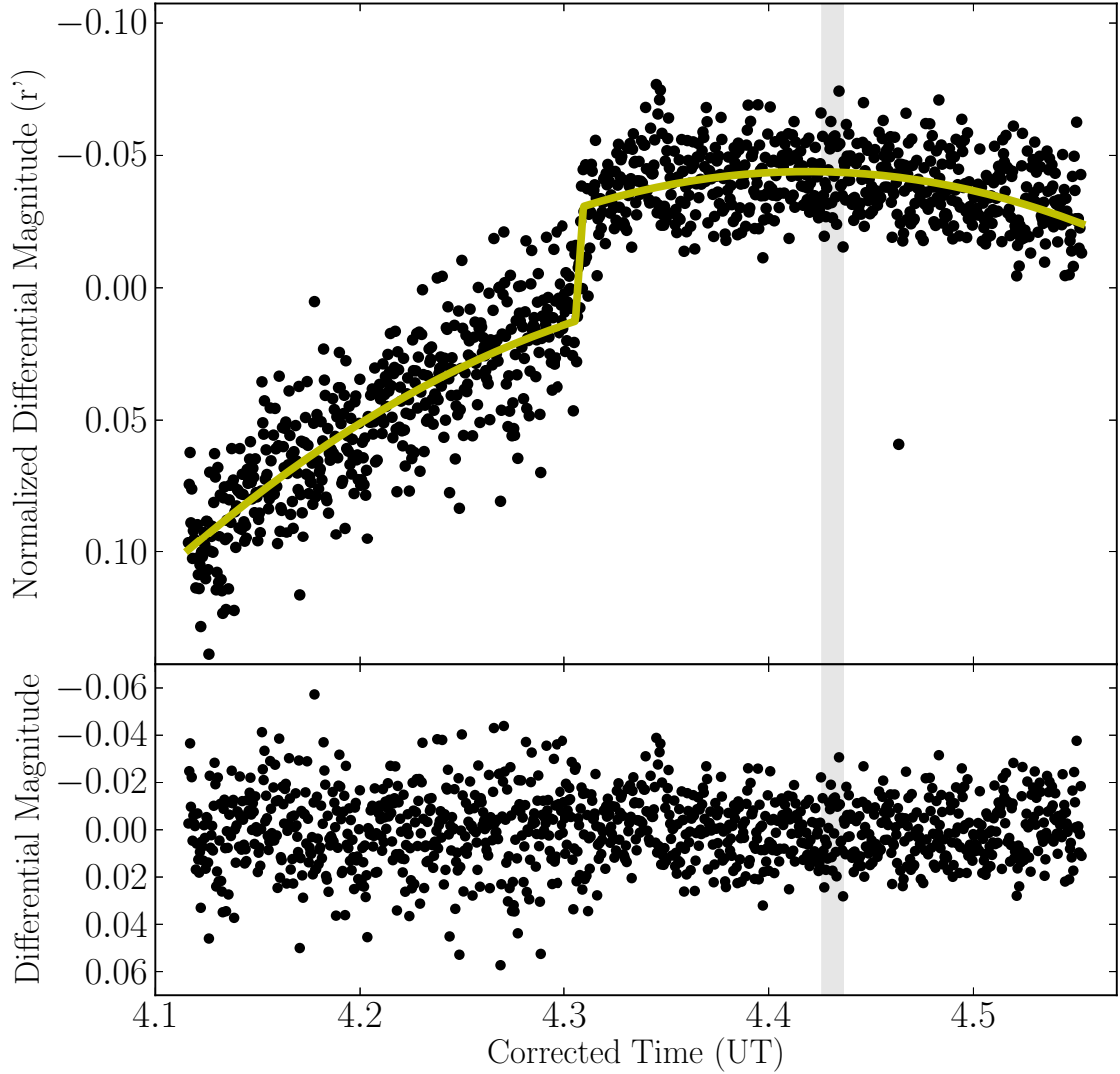} 
  \figcaption{\textbf{Top:} Normalized differential photometry. The best-fit photometric model, which has a parabolic component to account for the varying flux due to Quaoar's motion, and a linear offset to account for the artificial increase in brightness at $\sim4.3$~hrs (see Section~\ref{sec:observations}) is shown by the yellow curve. \textbf{Bottom:} The corrected differential photometry after removal of the best-fit model, which is sampled at 0.7~Hz and has a typical rms noise of 0.015 magnitude.  The time of closest approach as measured by the GMOS data is $04:25:52\pm7$~s, where any occultation signature is expected to occur, is shown by the grey shaded region.  \label{fig:photometry}}
\end{figure}

Twilight flats and bias frames were used to pre-calibrate the images using standard techniques. Relative photometry was measured using standard aperture photometry techniques; small equal sized apertures were centred on the combined Quaoar-target star image, and the two reference stars using the {\it IRAF} {\it center} routine of the {\it daophot} package. The {\it phot} task was used to measure the flux of all three sources inside those apertures. It was found that a 3.5 pixel radius aperture resulted in the lowest photometric shot noise. Relative variations in the combined flux of both reference stars compared to their mean summed flux were used to correct the flux of the Quaoar-target star for seeing and transparency variations. The resultant relative photometry is presented in Figure~\ref{fig:photometry}. 

The expectation of circular aperture photometry of a combined stationary and moving source is a peak in flux inside the aperture at time of smallest distance between the two sources, and flux that falls off parabolically with time before and after closest approach. The parabolic behaviour is apparent in the relative aperture photometry. In addition, the aperture photometry exhibits an increase in brightness of $\sim 0.05$ mags over a $\sim7$~s time interval at $\sim4.31$ hours UT. We attribute this to flat fielding errors in a region that becomes included in the aperture as the centroid, and hence the aperture, of the combined Quaoar+target star moves. Attempts to remove this feature through different flat fielding approaches, or by identifying and excluding the problematic region were unsuccessful. Rather, we chose to model the feature as a smooth linear transition of some amplitude that starts and stops at fixed times. We found that a 6 parameter model of a parabola (3 parameters) and the linear transition (3 parameters) could sufficiently describe the broad variations in observed photometry. This model was fit in a least-squares sense to the relative photometry, and removed. The residual corrected photometry, which is presented in Figure~\ref{fig:photometry}, is flat and is fully consistent with Gaussian noise with standard deviation of $0.015$ mags.

The photometry provides an independent measure of the time of closest approach; the best-fit parabola peaks at $04:25:13\pm27$~s UT, very close to the time of closest approach measured from the GMOS data, $04:25:52\pm7$. The error bars on each of these times is determined from the fits to the AC and GMOS observations (Fraser et al., in press). The small $\sim6$~s difference between the two times may be random fluctuation, or could be due to differences in the UT times recorded by the GMOS and AC header systems. 

\section{Atmospheric Limits}

To model any effects an atmosphere of Quaoar may have on the AC photometry, we make use of the model of \citet{Elliot1992}. We assume an isothermal atmosphere characterized by a single mean surface temperature and pressure. In addition, we assume the atmosphere is spherically symmetric. This assumption is justified by the observations of \citet{Ortiz2003} which reveal a 0.15 mag peak-to-peak variation in  Quaoar's lightcurve. If Quaoar's lightcurve variations are attributed to albedo variations, Quaoar's mean albedo cannot vary by more than $\sim15\%$ which would result in surface temperature variations of no more than $\sim3\%$ This temperature variation is much smaller than the expected latitudinal temperature change, and as such, the assumption of a spherically symmetric atmosphere seems justified. 

We consider model atmospheres with three different compositions of pure  CH$_4$, CO, and $N_2$ and assume that the atmosphere is sublimation supported. This fixes the atmospheric surface temperature given a mean surface pressure, or equivalently a pressure given a temperature. We adopt the pressure-temperature sublimation curves presented by \citet{Fray2009}. The model then has three parameters, surface pressure $p_{\textrm{s}}$, impact parameter $b$, and closest approach time, $T_{\textrm{CA}}$. 

For the properties of Quaoar, we adopt an apparent r' magnitude of 19.0 (Fraser et al., in press.), a mass of $1.34\times10^{21}$~kg from \citet{Fraser2013}, and a radius of $R=544\pm24$~km, the mean of the measurements of \citet{Braga-Ribas2013} and \citet{Fornasier2013}.

Thermal observations of Quaoar reveal a mean temperature of $\sim44$~K (Fornasier et al.; 2013, Mueller, T. personal communication). The exact temperature depends on Quaoar's unknown rotation pole angle, but is certainly no more than a few degrees from this value. For isothermal N$_2$ and CO, the surface pressure at Quaoar's temperatures would result in atmospheres with surface pressure of order a few microbars. These atmospheres would produce occultation signatures of depth $\gtrsim0.2$ magnitudes, even at 400~km above Quaoar's surface, the upper limit of the impact parameter measured from the GMOS data. This signature would be easily detected with our data. Thus, Quaoar cannot have a significant N$_2$ or CO atmosphere, the limits on which we quantify below.

To place limits on Quaoar's atmospheric extent, we turn to a Markov chain Monte Carlo (MCMC) likelihood approach. To probe likelihood space, we use {\it emcee}, an affine-invariant MCMC ensemble sampler \citep{Foreman-Mackey2012} and adopt a Gaussian likelihood of observing the photometry given a model atmosphere. The natural log-likelihood is then defined as

\begin{equation}
\log L=-\frac{1}{2}\sum_i \left(\frac{O_\textrm{i}-M_\textrm{i}(p_\textrm{s},b,T_{\textrm{CA}}, R)}{\sigma}\right)^2 
\label{eq:likelihood}
\end{equation}

\noindent
where $O_{\textrm{i}}$ are the observed relative magnitudes of the combined Quaoar+target star source, and $M_i(p_{\textrm{s}}, b,T_{\textrm{CA}})$ is the change in magnitude of the combined Quaoar (r'=19.0) target star (r'=20.2) as a result of refraction of stellar light by Quaoar's atmosphere. The free parameters are the surface pressure $p_{\textrm{s}}$, the impact parameter, $b$, the time of closest approach, $T_{\textrm{CA}}$, and Quaoar's radius, $R$. Finally, $\sigma$ is the photometric shot-noise.

Based on measurements from the GMOS data, we include a Gaussian prior on the impact parameter with mean 580~km and standard deviation of 120~km. In addition, we include a prior on time of closest approach with mean time, 4.43121 hours and standard deviation 6~s. We note that the prior on impact time has a very minimal effect on the results as a result of the nearly uniform photometric noise for all AC data. Finally, we consider a prior on Quaoar's radius. We adopt a uniform prior between 520 and 568 km, as is appropriate for the uncertainty in radius from \citet{Braga-Ribas2013} which is a full range on the radii they derive rather than a deviation.

In addition to the priors on $b$, $T_{\textrm{CA}}$, and $R$, we could have also considered a prior on the surface temperature which would essentially place a prior on $p_\textrm{s}$. We chose however, to forgo this prior, allowing our MCMC algorithm to place an independent upper limit on mean surface temperature, and leave comparison with thermal observations of Quaoar to the discussion.

In calculating $L$ for a given atmosphere model, we consider a window of width $w$ data points centred on the the close approach time, $T_{\textrm{CA}}$. In choosing $w$ one must consider that the atmosphere with the highest surface pressure allowable by the AC photometry will occur at the largest allowable impact parameter. Given Quaoar's temperature, occultation signatures at the largest allowable impact factor would produce detectable occultation signatures with width $\sim45-85$ seconds. The mean imaging frequency was 0.7~Hz, and as such we consider values of the window width, $32\leq w \leq 60$ frames. We note small variations in atmospheric constraints based on the choice of $w$, and choose to adopt the most conservative (highest pressure, temperature) limit found over the range of $w$ we consider. The photometric noise $\sigma$ was adopted as the standard deviation of 100 frames before and after the chosen window width.

To ensure accurate determination of the 3-$\sigma$ pressure upper limit, during the MCMC evaluation we used an ensemble of 200 walkers. A walker is an independent MCMC sampler that preserves its own posterior sample. Each walker underwent a burn-in phase of 200 steps which were discarded, and an additional 200 steps. The posterior distributions were estimated from the ensemble of post burn-in 200 steps of all walkers. Additional steps in either the burn-in or post burn-in phases did not noticeably change the posterior distributions, and we conclude that the MCMC algorithm converged.

\section{Results and Discussion}

\begin{figure}[h]
   \centering
   \plotone{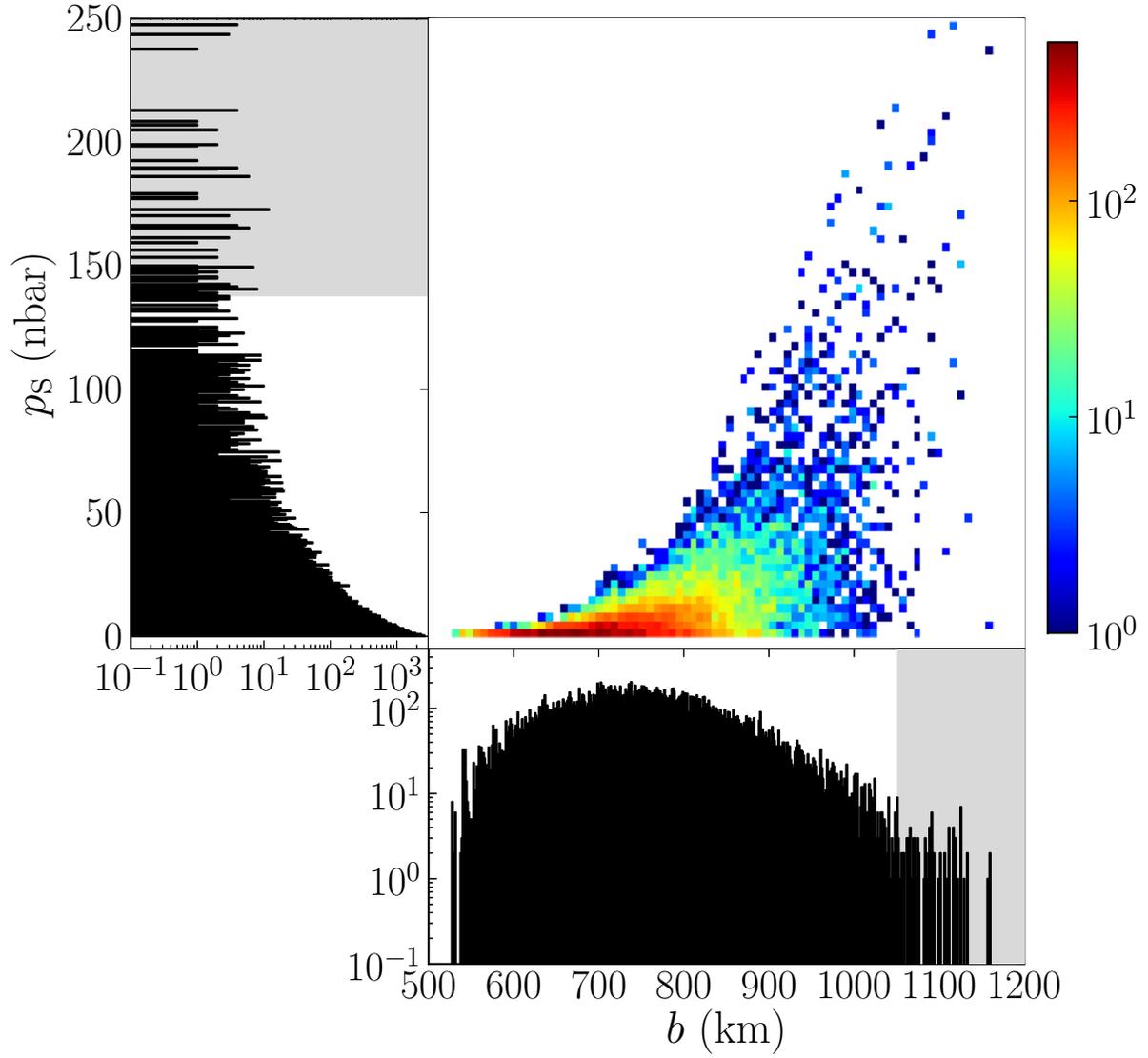} 
  \figcaption{Histograms of the posterior samples of surface pressure vs. impact parameter evaluated by the MCMC algorithm. Histograms are presented in log-space for clarity. Grey shaded regions show the region above the 3-$\sigma$ upper limit, and hence are excluded by the observations. \label{fig:posterior_CH4}}
\end{figure}

An example of the posterior distribution determined from our MCMC calculation is shown in Figure~\ref{fig:posterior_CH4} for a CH$_4$ atmosphere. Upper limits on the surface pressure were taken as the 99.7\% highest point in the sorted MCMC sample. As we assume Quaoar has an isothermal atmosphere that is in sublimation equilibrium with the surface ice, there is a 1-1 mapping between surface pressure and temperature. As such, the surface pressure upper limit is also an upper limit on surface temperature. From Figure~\ref{fig:posterior_CH4}, the 3-$\sigma$ surface pressure upper limit for the CH$_4$ atmosphere is 138~nbar, corresponding to a mean surface temperature of 44.3~K. This atmosphere corresponds to a pressure of 89~nbar at the 3-$\sigma$ impact parameter limit, or 365~km above Quaoar's surface, as measured by the GMOS data.

\citet{Braga-Ribas2013} present observations of a stellar occultation by Quaoar and do not detect an atmosphere. From those observations they place a surface pressure upper limit of 56~nbar for a methane dominated atmosphere. When determining this upper limit,  they assumed that Quaoar has a mean surface temperature of 42~K, and only allow the impact parameter to vary. As such, their upper limit does not fairly represent the effects of Quaoar's uncertain surface temperature. If the mean temperature is $44$~K rather than the $42$~K they assume, their upper limit surface pressure would be a factor of $\sim4$ higher, or $\sim200$ nbar, slightly larger than the the upper limit we find.

The 3-$\sigma$ upper limit on the surface pressure of an N$_2$ atmosphere is 5.3~$\mu\mbox{bar}$. At 365~km, the atmosphere would have a  pressure of 14~nbars. Assuming an isothermal sublimation supported atmosphere, this limit would require a mean surface temperature {\it upper limit} of $35.7$~K. This rules out the the existence of an N$_2$ atmosphere.

Quaoar's thermal flux is well described by a body of moderate thermal inertia, that has a sub-solar temperature of $\sim54$~K and mean temperature of $\sim44$~K \citep{Stansberry2008,Fornasier2013}. This temperature is significantly warmer than allowed by our observations; an isothermal N$_2$ atmosphere at Quaoar's temperature would result in more than an order of magnitude higher atmospheric pressure, and an occultation depth of at least 0.2 magnitudes. Such a warm N$_2$ atmosphere would produce a signal that would be easily detected in our data for occultations more than 1000~km off Quaoar's surface, a factor of 3 higher than the 3-$\sigma$ maximum allowable impact parameter. Uncertainties in the thermal flux, as well as uncertainties in Quaoar's pole position result in a few degree uncertainty in the mean surface temperature, but not enough to reconcile with the upper limit temperature of 33~K. Even if the assumption of a spherically symmetric atmosphere is removed, a N$_2$ atmosphere produced by a $\sim44$~K surface would have a microbar local surface temperature that would result in an easily detected asymmetric occultation signature. We conclude that Quaoar cannot have an N$_2$ atmosphere.

Given the similarity of the sublimation pressure-temperature behaviour of CO and N$_2$ ices, we can draw the same conclusions about CO ice. That is, the 3-$\sigma$ upper limit of the surface temperature and pressure of an isothermal CO atmosphere is 2.7 $\mu\mbox{bar}$ and 38~K. Like for an N$_2$ atmosphere, this temperature is much too cold to be consistent with the thermal observations, and we conclude that Quaoar cannot have an isothermal CO atmosphere.

The upper limit  mean temperature of the CH$_4$ atmosphere, 44.3~K, is very similar to the mean temperature inferred from thermal observations of the body. Thus, unlike CO and N$_2$, we cannot eliminate the possibility that Quaoar has an isothermal CH$_4$ atmosphere.  This is in agreement with the observations of \citet{Schaller2007b} that reveal the signatures of methane absorption in Quaoar's reflectance spectrum. It seems likely that Quaoar does bear a methane-dominated atmosphere, but with a mean temperature and pressure just beyond the limits of detectability from the observations we present here.

\acknowledgements
The authors thank the Director of the Gemini telescopes for time awarded under the director's discretion.

Based on observations obtained at the Gemini Observatory, which is operated by the Association of Universities for Research in Astronomy, Inc., under a cooperative agreement with the NSF on behalf of the Gemini partnership: the National Science Foundation (United States), the National Research Council (Canada), CONICYT (Chile), the Australian Research Council (Australia), Minist\'{e}rio da Ci\^{e}ncia, Tecnologia e Inova\c{c}\~{a}o  (Brazil) and Ministerio de Ciencia, Tecnolog\'{i}a e Innovaci\'{o}n Productiva (Argentina).


\begin{figure}[h]
   \centering
   \plotone{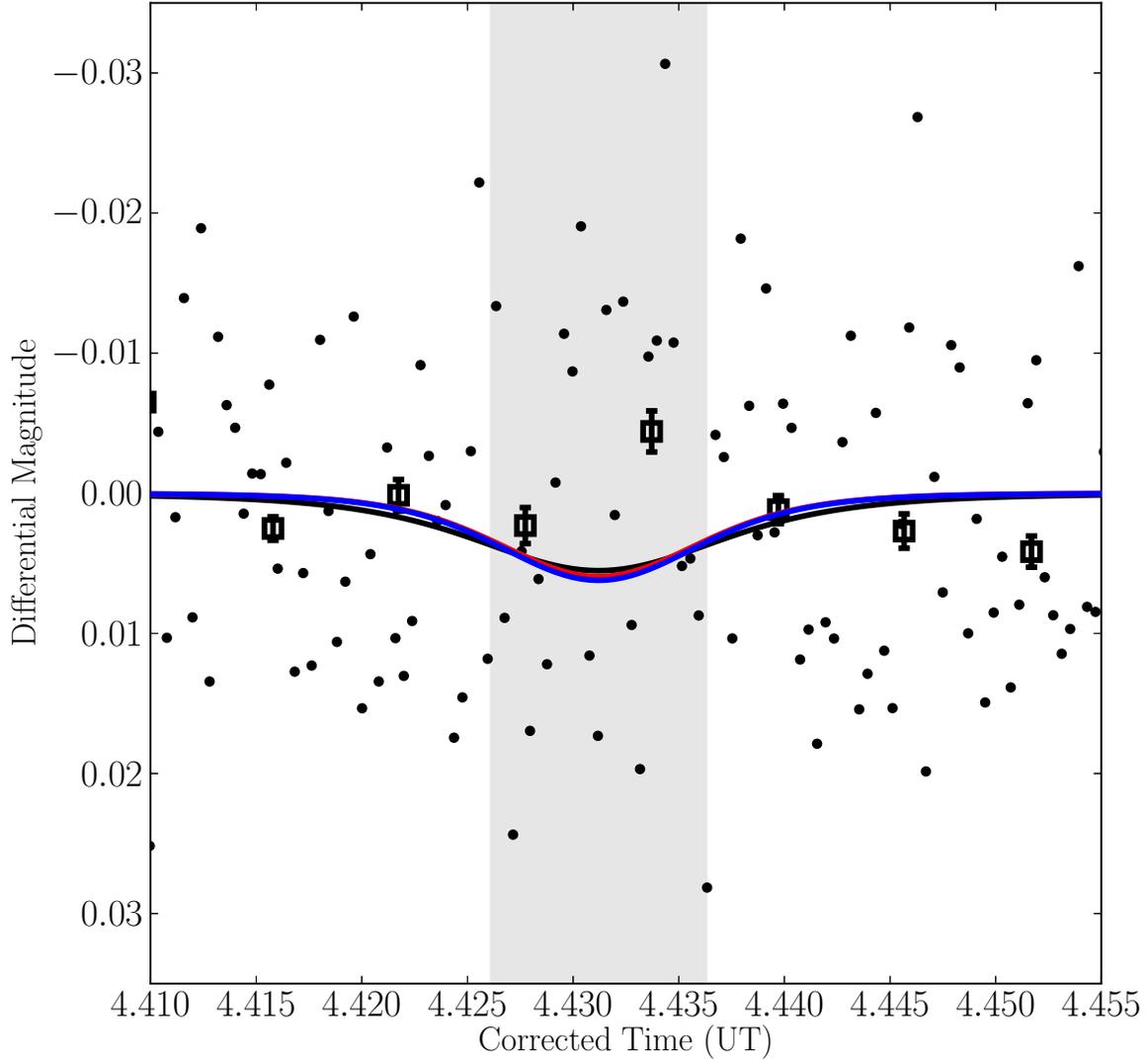} 
  \figcaption{Corrected differential photometry (black points) and photometry binned by a factor of 15 (open black squares). Errorbars represent the binned photometric uncertainty and are typically 4 millimags. The 3-$\sigma$ upper limit CH$_4$, N$_2$, and CO atmospheres excluded by the corrected photometry are shown in black, red, and blue respectively. For reference, the time of closest approach, which is shown by the grey shaded region, is $04:25:52\pm7$~s UT.  \label{fig:occult_fig}}
\end{figure}

\end{document}